\documentclass[12pt,preprint]{aastex}
\usepackage{epsf,graphics}
\usepackage[dvips,usenames]{color}

\begin{document}
\begin{flushright}
 Nature, 426, 810-812 (2003)
\end{flushright}
{\bf
 A gravitationally lensed quasar with quadruple images separated \\
by 14.62 arcseconds 
}
\bigskip\\
Naohisa Inada$^1$,
Masamune Oguri$^1$,
Bartosz Pindor$^2$,
Joseph F. Hennawi$^2$,\\
Kuenley Chiu$^3$,
Wei Zheng$^3$,
Shin-Ichi Ichikawa$^4$,
Michael D. Gregg$^{5,6}$,\\
Robert H. Becker$^{5,6}$,
Yasushi Suto$^1$,
Michael A. Strauss$^2$,
Edwin L. Turner$^2$,\\
Charles R. Keeton$^7$,
James Annis$^8$,
Francisco J. Castander$^9$,
Daniel J. Eisenstein$^{10}$,\\
Joshua A. Frieman$^{7,8}$,
Masataka Fukugita$^{11}$,
James E. Gunn$^2$,
David E. Johnston$^7$,\\
Stephen M. Kent$^8$,
Robert C. Nichol$^{12}$,
Gordon T. Richards$^2$,
Hans-Walter Rix$^{13}$,\\
Erin Scott Sheldon$^7$,
Neta A. Bahcall$^2$,
J. Brinkmann$^{14}$,
\v{Z}eljko Ivezi\'{c}$^2$,\\
Don Q. Lamb$^7$,
Timothy A. McKay$^{15}$,
Donald P. Schneider$^{16}$
\&
Donald G. York$^{7,17}$
\bigskip\\
{\footnotesize
{}$^1$\,{Department of Physics, School of Science,
University of Tokyo, 113-0033, Japan}\\
{}$^2$\,{Princeton University Observatory, Peyton Hall,
Princeton, NJ 08544, USA}\\
{}$^3$\,{Department of Physics and Astronomy, Johns Hopkins
University, 3701 San Martin Drive, Baltimore, \\
 \hspace*{5mm}MD 21218, USA}\\
{}$^4$\,{National Astronomical Observatory, 2-21-1 Osawa,
Mitaka, Tokyo 181-8588, Japan}\\
{}$^5$\,{Department of Physics, University of California at
Davis, 1 Shields Avenue, Davis, CA 95616, USA}\\
{}$^6$\,{Institute of Geophysics and Planetary Physics,
Lawrence Livermore National Laboratory, L-413, \\
\hspace*{5mm}7000 East Aveneu, Livermore, CA 94550, USA}\\
{}$^7$\,{Department of Astronomy and Astrophysics,
University of Chicago, 5640 South Ellis Avenue, Chicago, \\
 \hspace*{5mm}IL 60637, USA}\\
{}$^8$\,{Fermi National Accelerator Laboratory, P.O. Box
500, Batavia, IL 60510, USA}\\
{}$^9$\,{Institut d'Estudis Espacials de Catalunya/CSIC,
Gran Capita 2-4, 08034 Barcelona, Spain}\\
{}$^{10}$\,{Steward Observatory, University of Arizona,
933 North Cherry Avenue, Tucson, AZ 85721, USA}\\
{}$^{11}$\,{Institute for Cosmic Ray Research, University of Tokyo,
5-1-5 Kashiwa, Kashiwa City, \\
 \hspace*{5mm}Chiba 277-8582, Japan}\\
{}$^{12}$\,{Department of Physics, Carnegie Mellon University,
Pittsburgh, PA 15213, USA}\\
{}$^{13}$\,{Max-Planck Institute for Astronomy, K\"onigstuhl
17, D-69117 Heidelberg, Germany}\\
{}$^{14}$\,{Apache Point Observatory, P.O. Box 59, Sunspot, NM88349, USA}\\
{}$^{15}$\,{Department of Physics, University of Michigan, 500 East
University Avenue, Ann Arbor, MI 48109, USA}\\
{}$^{16}$\,{Department of Astronomy and Astrophysics,
Pennsylvania State University, 525 Davey Laboratory, \\
 \hspace*{5mm}University Park, PA 16802, USA}\\
{}$^{17}$\,{Enrico Fermi Institute, University of Chicago, 5640 South
Ellis Avenue, Chicago, IL 60637, USA}
}
\bigskip

{\bf
Gravitational lensing is a powerful tool for the study of the
distribution of dark matter in the Universe. The cold-dark-matter model
of the formation of large-scale structures predicts$^{1-6}$ the
existence of quasars gravitationally lensed by concentrations of dark
matter$^{7}$ so massive that the quasar images would be split by over 7
arcsec. Numerous searches$^{8-11}$ for large-separation lensed quasars
have, however, been unsuccessful. All of the roughly 70 lensed quasars
known$^{12}$, including the first lensed quasar discovered$^{13}$, have
smaller separations that can be explained in terms of galaxy-scale
concentrations of baryonic matter. Although gravitationally lensed
galaxies$^{14}$ with large separations are known, quasars are more
useful cosmological probes because of the simplicity of the resulting
lens systems. Here we report the discovery of a lensed quasar, SDSS
J1004+4112, which has a maximum separation between the components of
14.62 arcsec. Such a large separation means that the lensing object must
be dominated by dark matter. Our results are fully consistent with
theoretical expectations$^{3-5}$ based on the cold-dark-matter model. 
}

For bright quasars ($i$-band magnitude $i<19$), the probability of
gravitational lensing$^{15}$ is only about 0.1\%; the majority of these
lenses have small separations, due to a single massive galaxy. The
fraction of large-separation lensed quasars is predicted to be 0.01\% or
less$^{3-5}$; thus it is not surprising that none have been found to
date$^{8-11}$. In order to find such objects, we need samples of tens of
thousands of quasars, such as generated by the Sloan Digital Sky
Survey$^{16,17}$ (SDSS).  The SDSS is conducting both a photometric
survey$^{18-22}$ using five broad optical bands$^{22}$ ($u$, $g$, $r$,
$i$ and $z$) and a spectroscopic survey$^{23}$ of 10,000 square degrees
of the sky centered approximately on the North Galactic Pole, using a
dedicated wide-field 2.5-m telescope at the Apache Point Observatory.

We searched for large-separation lensed quasars in a sample of
$\sim$29,500 spectroscopically-confirmed SDSS quasars$^{24}$ at
redshifts $z$ of $0.6-2.3$, a sample roughly three times larger than
those used in previous searches.  Even with this large sample, the
expected number of large-separation lensed quasars is of the order of
unity.  In the field around each quasar in the sample, we searched for
stellar objects with colours differing by less than 0.1 from those of
the quasar, with separations 
between $7\farcs0$ and $60\farcs0$ and with flux greater than one-tenth
that of the quasar. SDSS~J1004+4112 was identified as a `quadruple'
large-separation lensed quasar candidate using these criteria.  Only one
of the four components (component B, see below) has an SDSS spectrum
(the SDSS hardware$^{23}$ does not allow pairs of objects separated by
less than $55''$ to be observed on a single plate), and therefore, we
obtained spectra of all four components using the Keck~I telescope at
the W. M. Keck Observatory.  The results are shown in Fig. 1.  All
four components indeed show quasar-like features, with all emission
lines giving a consistent redshift $z=1.734{\pm}0.002$; the velocity
differences of the quasar components are $\sim 100$ km s$^{-1}$,
comparable to the observational uncertainty. Although it is not obvious
from Fig. 1, there are \ion{C}{4} absorption line systems at $z=1.732$
in each of the four quasar spectra; this is an absorption system
associated with the quasar itself, further supporting the lensing
hypothesis: the four quasar images are from the same physical source.
The differences in their spectra may be explained by the modest
time-variability of the source quasar over ${\sim}1$ year, the expected
gravitational lensing time-delay$^{25}$ among those different images.

Additional strong support for the lensing hypothesis comes from the
identification of the galaxy cluster responsible for the
large-separation lensing. From the observed image separations
(the maximum separation is $14\farcs62$, between images B and C), we
infer that the lensing object should have a velocity dispersion in
excess of 600 km s${}^{-1}$.  Thus the lensing object cannot be a single
galaxy, but must be rather a group or cluster of galaxies that has a
sufficiently concentrated distribution of dark matter. To identify the
lensing object, we obtained deep optical images of the system using the
Subaru telescope of the National Astronomical Observatory, Japan. The
result is shown in Fig. 2. A number of galaxies are clearly detected
around component G, suggesting that it is the most luminous galaxy of
the cluster. We obtained a spectrum of component G using the Keck I
telescope. The spectrum shows a number of absorption features
characteristic of a early-type galaxy at a redshift of
$z=0.6799\pm0.0001$. We also obtained spectra of two faint galaxies
immediately to the south-west of component G (Fig. 2) using the Faint
Object Camera and Spectrograph$^{26}$ of the Subaru telescope. The
redshifts of these two faint galaxies are $z=0.6751\pm0.0001$, strongly
suggesting a cluster of galaxies at $z{\sim}0.68$ centred on component
G. Clusters are dominated by elliptical galaxies, which all have very
similar spectral energy distributions. 

Many of the faint galaxies in Fig. 2 (${\sim}40$ galaxies around
component G) have colours that are 
similar to that of component G. The colours are consistent with the
expected colours of elliptical galaxies at $z{\sim}0.68$ ($g-r{\sim}1.8$
and $r-i{\sim}1.1$). In addition, there is an X-ray source in this
direction detected by the ROSAT All-Sky Survey$^{27}$ (0.236 counts per
second in a 473-s exposure). The emission, however, comes most
probably from the quasar, because the detected X-ray flux is too strong
for typical clusters of galaxies at $z=0.68$. Finally, we note two
possible arclets (highly distorted images of background galaxies due to
gravitational lensing) in Fig. 2 (marked as `arc?') close to
component D. If future observations confirm that the arclets are indeed
lensed background galaxies, they will provide strong additional
constraints on the total mass distribution of the lensing cluster.

The lensing interpretation is further supported by a theoretical model
of SDSS~J1004+4112.  We fitted the positions of the four quasar components
with a singular isothermal ellipsoid (SIE) plus external shear model
using lens modeling software$^{28}$. The best-fit model is
illustrated in Fig. 3. The positions and relative brightnesses of all
components agree well with the lens model predictions. The centre of the
lensing mass is offset from the centre of component G by about 10 kpc at
the cluster redshift, but brightest galaxy of a cluster is not always
found exactly at the centre of the potential well of that cluster. 

The identical redshifts ($z=1.734$) and the spectral energy
distributions of the four lensed components, the existence of a lensing
cluster of galaxies ($z=0.68$), and the presence of possible
arclets confirm the hypothesis that the quasar is lensed by this
cluster. Furthermore, a theoretical lensing model involving the cluster
and external shear simultaneously accounts for the observed geometry of
the system and the relative brightness of the images. The present work
represents the discovery of a long-predicted but previously undetected
population of large-separation lensed quasars. 

\bigskip

\noindent
{\footnotesize
\begin{enumerate}
\item Narayan, R. \& White, S. D. M.
   Gravitational lensing in a cold dark matter universe.
   {\sl Mon. Not. R. Astron. Soc.} {\bf 231}, 97--103 (1988).
\item Wambsganss, J., Cen, R., Ostriker, J. P. \& Turner, E. L.
     Testing Cosmogonic Models with Gravitational Lensing.
     {\sl Science} {\bf 268}, 274--276 (1995).
\item Keeton, C. R. \& Madau, P.
   Lensing Constraints on the Cores of Massive Dark Matter Halos.
   {\sl Astrophys. J.} {\bf 549}, L25--L28 (2001).
\item Wyithe, J. S. B., Turner, E. L. \& Spergel, D. N.
   Gravitational Lens Statistics for Generalized NFW Profiles: Parameter
   Degeneracy and Implications for Self-Interacting Cold Dark Matter.
   {\sl Astrophys. J.} {\bf 555}, 504--523 (2001).
\item Takahashi, R. \& Chiba, T.
     Gravitational Lens Statistics and the Density Profile of Dark Halos.
     {\sl Astrophys. J.} {\bf 563}, 489--496 (2001).
\item Oguri, M.
    Constraints on the Baryonic Compression and Implications for the
    Fraction of Dark Halo Lenses.
    {\sl Astrophys. J.} {\bf 580}, 2--11 (2002).
\item Navarro, J. F., Frenk, C. S. \& White, S. D. M.
    A Universal Density Profile from Hierarchical Clustering.
    {\sl Astrophys. J.} {\bf 490}, 493--508 (1997).
\item Maoz, D., Rix, H., Gal-Yam, A. \& Gould, A.
   Survey for Large--Image Separation Lensed Quasars.
   {\sl Astrophys. J.} {\bf 486}, 75--84 (1997).
\item Ofek, E. O., Maoz, D., Prada, F., Kolatt, T. \& Rix, H.
   A survey for large-separation lensed FIRST quasars.
   {\sl Mon. Not. R. Astron. Soc.} {\bf 324}, 463--472 (2001).
\item Phillips, P. M. {\it et al.}
   The JVAS/CLASS search for 6-arcsec to 15-arcsec image separation
   lensing.
   {\sl Mon. Not. R. Astron. Soc.} {\bf 328}, 1001--1015 (2001).
\item Zhdanov, V. I. \& Surdej, J.
     Quasar pairs with arcminute angular separations.
     {\sl Astron. \& Astrophys.} {\bf 372}, 1--7 (2001).
\item Kochanek, C. S. {\it et al.}
     CASTLES Survey. at $<$http://cfa-www.harvard.edu/castles/$>$
     (2003). 
\item Walsh, D., Carswell, R. F. \& Weymann, R. J.
     0957 + 561 A, B - Twin quasistellar objects or gravitational lens.
     {\sl Nature} {\bf 279}, 381--384 (1979).
\item Colley, W. N., Tyson, J. A. \& Turner, E. L.
    Unlensing Multiple Arcs in 0024+1654: Reconstruction of the Source
    Image. 
    {\sl Astrophys. J.} {\bf 461}, L83--L86 (1996).
\item Turner, E. L., Ostriker, J. P. \& Gott, J. R., III
     The statistics of gravitational lenses -- The distributions of
     image angular separations and lens redshifts.
     {\sl Astrophys. J.} {\bf 284}, 1--22 (1984).
\item York, D. G. {\it et al.}
   The Sloan Digital Sky Survey: Technical Summary.
   {\sl Astron. J.} {\bf 120}, 1579--1587 (2000).
\item Stoughton, C. {\it et al.}
    Sloan Digital Sky Survey: Early Data Release.
    {\sl Astron. J.} {\bf 123}, 485--548 (2002).
\item Gunn, J. E. {\it et al.}
    The Sloan Digital Sky Survey Photometric Camera.
    {\sl Astron. J.} {\bf 116}, 3040--3081 (1998).
\item Pier J. R. {\it et al.}
    Astrometric Calibration of the Sloan Digital Sky Survey.
    {\sl Astron. J.} {\bf 125}, 1559--1579 (2003).
\item Hogg, D. W., Finkbeiner, D. P., Schlegel, D. J. \& Gunn, J. E.
    A Photometricity and Extinction Monitor at the Apache Point
    Observatory. 
    {\sl Astron. J.} {\bf 122}, 2129--2138 (2001).
\item Smith, J. A. {\it et al.}
    The $u'g'r'i'z'$ Standard-Star System.
    {\sl Astron. J.} {\bf 123}, 2121--2144 (2002).
\item Fukugita, M. {\it et al.}
   The Sloan Digital Sky Survey Photometric System.
   {\sl Astron. J.} {\bf 111}, 1748--1756 (1996).
\item Blanton, M. R. {\it et al.}
    An Efficient Targeting Strategy for Multiobject Spectrograph
    Surveys: the Sloan Digital Sky Survey ``Tiling'' Algorithm.
    {\sl Astron. J.} {\bf 125}, 2276--2286 (2003).
\item Richards, G. T. {\it et al.}
   Spectroscopic Target Selection in the Sloan Digital Sky Survey:
   The Quasar Sample.
   {\sl Astron. J.} {\bf 123}, 2945--2975 (2002).
\item Oguri, M., Taruya, A., Suto, Y. \& Turner, E. L.
    Strong Gravitational Lensing Time Delay Statistics and the Density
    Profile of Dark Halos.
    {\sl Astrophys. J.} {\bf 568}, 488--499 (2002).
\item Kashikawa, N. {\it et al.}
    FOCAS: The Faint Object Camera and Spectrograph for the Subaru
    Telescope. 
    {\sl Publications of the Astronomical Society of Japan} 
    {\bf 54}, 819--832 (2002).
\item Cao, L., Wei, J.-Y. \& Hu, J.-Y.
     High X-ray-to-optical flux ratio RASS-BSC sources. I. The optical
     identification. 
     {\sl Astron. \& Astrophys. Supplement} {\bf 135}, 243--253 (1999).
\item Keeton, C. R.
    Computational Methods for Gravitational Lensing.
    Preprint astro-ph/0102340 at $<$http://xxx.lanl.gov$>$ (2001).
\item Oke, J. B. {\it et al.}
    The Keck Low-Resolution Imaging Spectrometer.
    {\sl Publications of the Astronomical Society of the Pacific} 
    {\bf 107}, 375--385 (1995).
\item Miyazaki, S. {\it et al.}
    Subaru Prime Focus Camera -- Suprime-Cam.
    {\sl Publications of the Astronomical Society of Japan} 
    {\bf 54}, 833--853 (2002).
\end{enumerate}
}

\bigskip

\noindent
{\bf Acknowledgments} Funding for the creation and distribution of the
SDSS Archive has been provided by the Alfred P. Sloan Foundation, the
Participating Institutions, the National Aeronautics and Space
Administration, the National Science Foundation, the U.S. Department of
Energy, the Japanese Monbukagakusho, and the Max Planck Society. The
SDSS Web site is http://www.sdss.org/.

The SDSS is managed by the Astrophysical Research Consortium (ARC) for
the Participating Institutions. The Participating Institutions are The
University of Chicago, Fermilab, the Institute for Advanced Study, the
Japan Participation Group, The Johns Hopkins University, Los Alamos
National Laboratory, the Max-Planck-Institute for Astronomy (MPIA), the
Max-Planck-Institute for Astrophysics (MPA), New Mexico State
University, University of Pittsburgh, Princeton University, the United
States Naval Observatory, and the University of Washington.  

This paper is based in part on data collected at the Subaru Telescope,
which is operated by the National Astronomical Observatory of Japan,
W. M. Keck Observatory, which is operated as a scientific partnership
among the California Institute of Technology, the University of
California, and the National Aeronautics and Space Administration, and
the Apache Point Observatory (APO) 3.5-meter telescope, which is owned
and operated by the Astrophysical Research Consortium.

Part of this work was performed under the auspices of the U.S.
Department of Energy at the University of California Lawrence Livermore
National Laboratory. 

\bigskip

\noindent
{\bf Competing interests statement} The authors declare that they have no
competing financial interests.

\bigskip

\noindent
{\bf Correspondence} and requests for materials should be addressed to
N.I. \\ (e-mail: inada@utap.phys.s.u-tokyo.ac.jp) 

\clearpage
\begin{deluxetable}{crrccccr}
\tabletypesize{\small}
\tablecolumns{8}
\tablewidth{0pc}
\tablecaption{{\bf ASTROMETRY AND PHOTOMETRY FOR SDSS~J1004+4112}}
\tablehead{
\colhead{Object} & \colhead{R.A.(J2000)\tablenotemark{a}} & 
\colhead{Dec.(J2000)\tablenotemark{a}} &
\colhead{$g$\tablenotemark{b}} & \colhead{$r$\tablenotemark{b}} & \colhead{$i$\tablenotemark{b}}
& \colhead{$z$\tablenotemark{b}} & \colhead{$\Delta{\theta}$\tablenotemark{c}}}
\startdata
A & 10 04 34.794 & +41 12 39.29 & 18.67$\pm$0.03 & 18.70$\pm$0.02 & 18.46$\pm$0.02 & 18.43$\pm$0.05 &  $3\farcs73$  \\
B & 10 04 34.910 & +41 12 42.79 & 19.05$\pm$0.06 & 19.10$\pm$0.06 & 18.86$\pm$0.06 & 18.92$\pm$0.06 &  $0\farcs00$   \\
C & 10 04 33.823 & +41 12 34.82 & 19.71$\pm$0.03 & 19.73$\pm$0.02 & 19.36$\pm$0.03 & 19.31$\pm$0.07 &  $14\farcs62$   \\
D & 10 04 34.056 & +41 12 48.95 & 20.67$\pm$0.04 & 20.51$\pm$0.04 & 20.05$\pm$0.04 & 20.00$\pm$0.13 &  $11\farcs44$  \\
G & 10 04 34.170 & +41 12 43.66 & 22.11$\pm$0.40 & 20.51$\pm$0.13 & 19.54$\pm$0.09 & 19.04$\pm$0.21 &  $ 8\farcs44$  \\
\enddata
\tablenotetext{a}{R.A., right ascension; Dec., declination. RA and Dec.
 are given in the J2000 system. These celestial
 coordinates were measured on the basis of the celestial coordinates of
 component B. The positional errors of components A, C, and D (not
 including the absolute positional errors of component B) are
 $0\farcs01$ and that of component G is $0\farcs05$ per coordinate.} 
\tablenotetext{b}{$g$, $r$, $i$ and $z$ mean the magnitudes of each
 band.} 
\tablenotetext{c}{Separation angles relative to component B}
\end{deluxetable}

\clearpage

\begin{figure}[t]
\epsscale{0.65}
\plotone{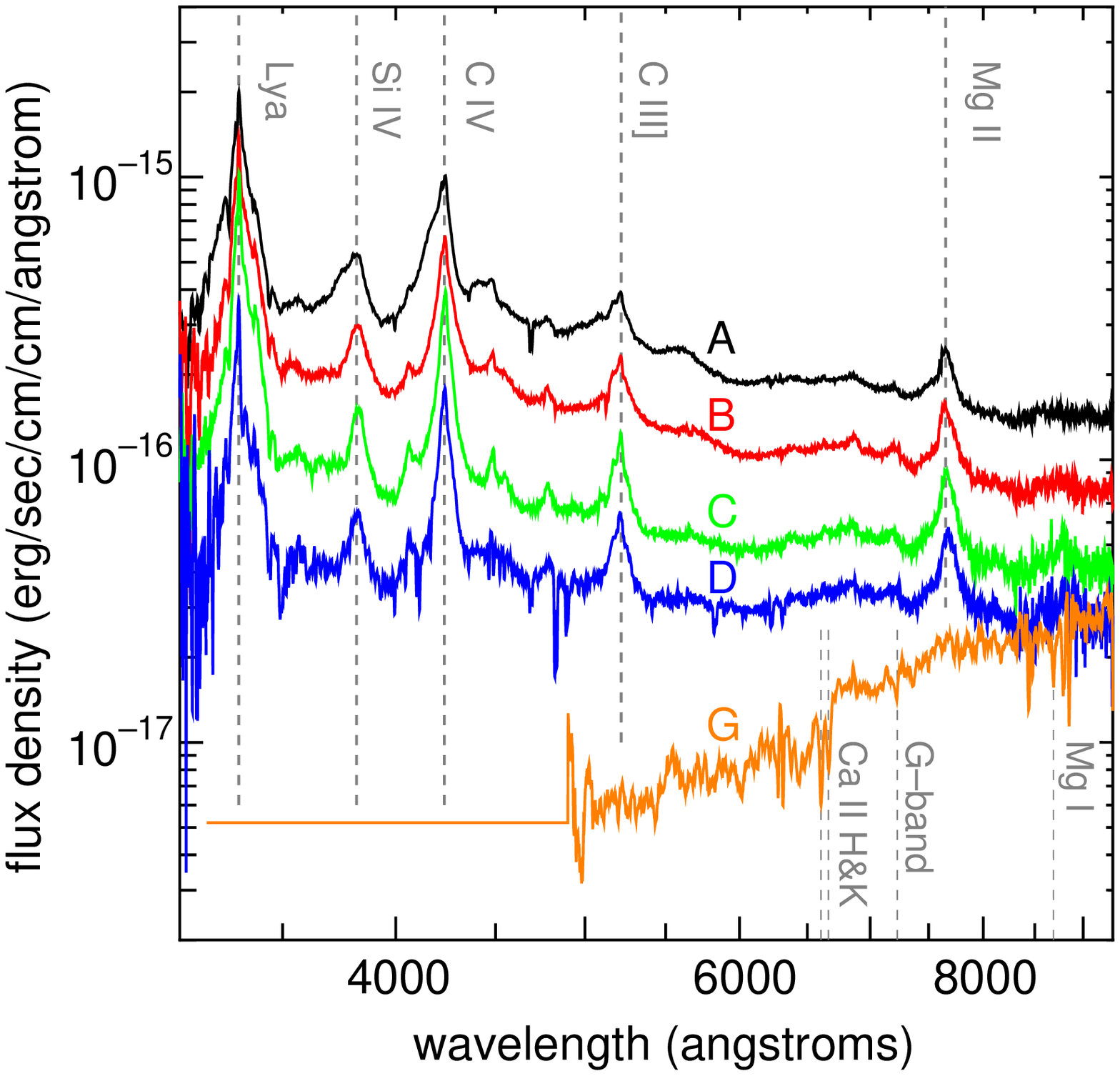}
\end{figure}


\noindent
{\small
{\bf Figure 1} The Keck spectra of the four quasar components A--D 
and the brightest galaxy G in the lensing cluster. See Fig. 2 for
these identifications (A--D, and G). The data were taken using the
Low-Resolution Imaging Spectrometer$^{29}$ (LRIS) of the Keck~I
telescope. The exposure times were 900 s for each component. The
dispersion is 1.09{\AA} pixel$^{-1}$. The data were reduced in a
standard method using IRAF (IRAF is the image reduction and analysis
facility, distributed by the National Optical Astronomy Observatories,
which are operated by the Association of Universities for Research in
Astronomy, Inc., under cooperative agreement with the National Science
Foundation). The black solid line, the red solid line, the green solid
line, and the blue solid line represent the spectra of components A, B,
C, and D, respectively. The vertical gray dotted lines (3323.6 {\AA},
3818.7 {\AA}, 4235.1 {\AA}, 5218.5 {\AA}, and 7651.8{\AA}) represent the
positions of emission lines of the respective ions redshifted to
$z=1.734$ of Ly$\alpha$ (1215.67 {\AA}), \ion{Si}{4} (1396.76 {\AA}),
\ion{C}{4} (1549.06 {\AA}), \ion{C}{3]} (1908.73 {\AA}), and \ion{Mg}{2}
(2798.75 {\AA}), respectively. All emission lines are clearly at the
same redshift. The orange solid line represents the Keck spectrum of
component G at the same dispersion.  The exposure time was also 900
s for component G. The vertical thinner gray dotted lines (3933.7
{\AA}, 3968.5 {\AA}, 4304.4 {\AA}, and 5175.3 {\AA}) represent the
positions of absorption lines of the respective ions redshifted to 
$z=0.680$ of \ion{Ca}{2} H\&K (3933.7 {\AA} and 3968.5 {\AA}), G-band 
(4304.4 {\AA}), and \ion{Mg}{1} b-band (5175.3 {\AA}), respectively.
There are no data below $\sim4900${\AA} in the spectrum of component G.  
}

\clearpage

\begin{figure}[t]
\epsscale{0.9}
\plotone{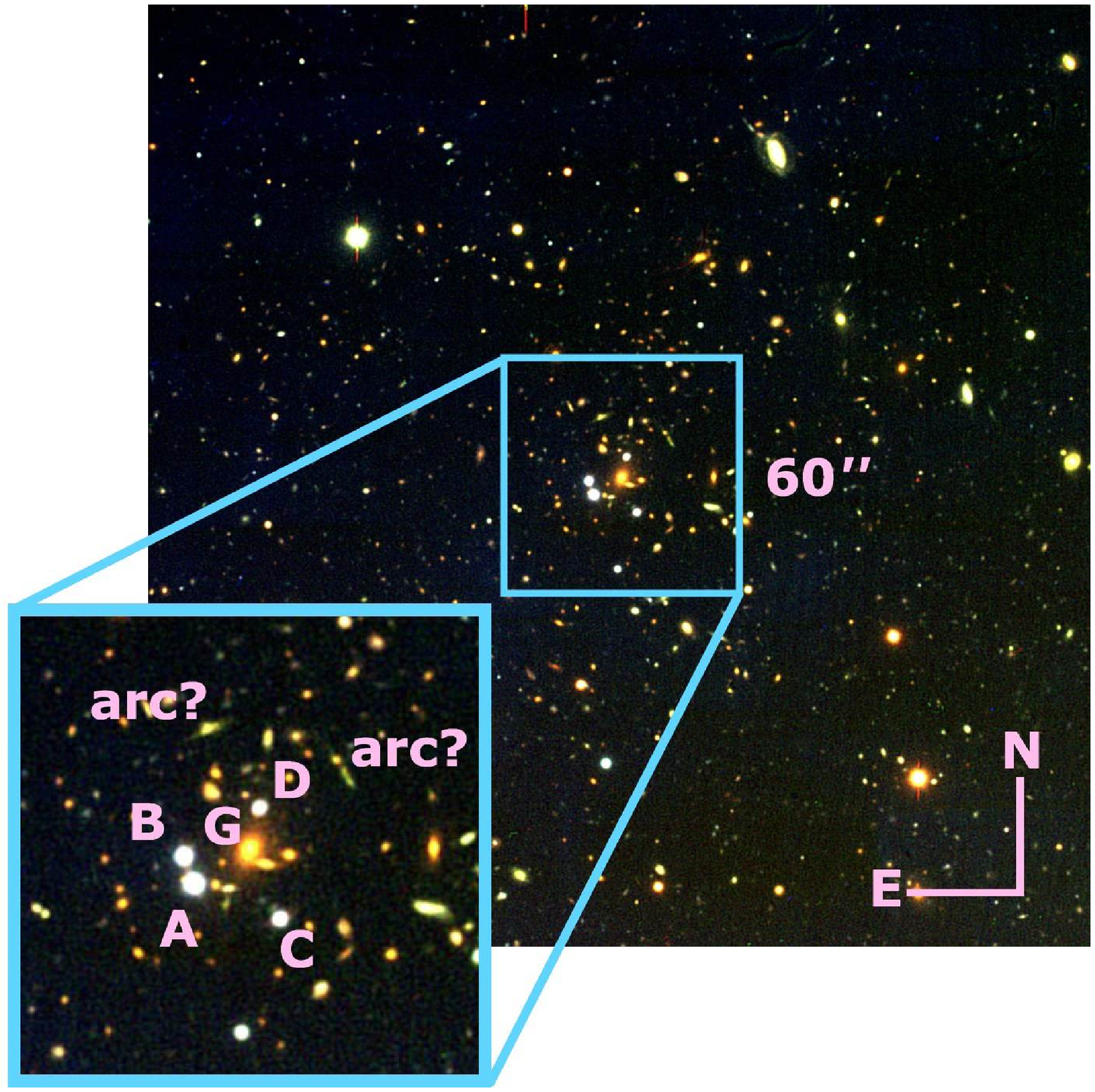}
\end{figure}


\noindent
{\small
{\bf Figure 2} The $gri$ composite Subaru image of the field around SDSS
J1004+4112. The data were taken using the Subaru Prime Focus
Camera$^{30}$ of the Subaru telescope. The magnitude limit is ${i}
\approx 26.0$. The central $60''$ square is shown in an expanded view.
The four quasar components are marked as A, B, C and D, and the bright
galaxy located between the four quasar components is marked as G. The
separation between components A and D is $12\farcs77$, and that between
components B and C is $14\farcs62$. The positions (J2000) and the
magnitudes of the components A--D and the brightest galaxy (component G)
between the four quasar components are summarized in Table~1. Many faint
galaxies can be seen --- their positions and colours are consistent with
being members of a cluster ($z=0.68$) centred on component G. Two
possible arclets (marked as `arc?') can also be seen. The seeing had a
full-width at half-maximum of $0\farcs6$. 
}

\clearpage

\begin{figure}[t]
\epsscale{0.6}
\plotone{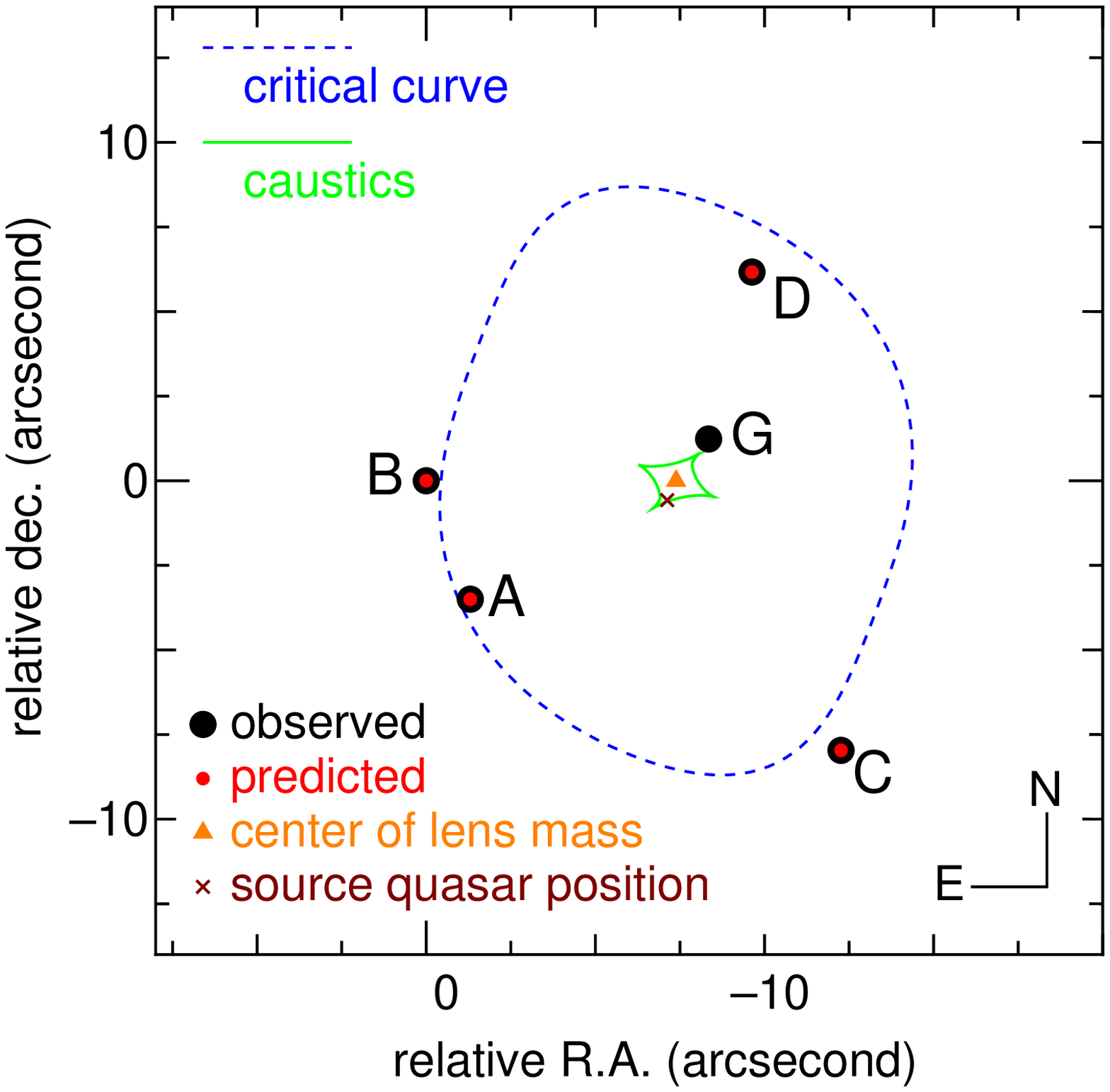}
\end{figure}


\noindent
{\small
{\bf Figure 3} The best fit lens model prediction compared with the
observation.  We used a lensing model of a singular isothermal ellipsoid
(SIE) with external shear.  The best-fit model has an Einstein radius of the
SIE model ${{\alpha}_{e}}=6\farcs906$ (corresponding to a velocity
dispersion $\sim700$ km sec$^{-1}$ at the cluster redshift), magnitude
and position angle of the shear $\gamma=0.250$ and
$\theta_{\gamma}=-60.925^\circ$ (measured East of North), and
ellipticity and its position angle $e=0.498$ and
${\theta}_{e}=21.434^\circ$ (measured East of North), with a source
quasar position ($\Delta$R.A., $\Delta$Dec.)$=$($-7\farcs124$,
$-0\farcs574$) and a centre of lensing mass ($\Delta$R.A.,
$\Delta$Dec.)$=$($-7\farcs387$, $-0\farcs004$) relative to the centre of
component A. The black filled circles represent the observed positions
of components A, B, C, D and G, and the red filled circles represent
the predicted positions of components A--D. The green solid line is the
position of the caustic in the source plane, and the blue dashed line
represent the critical curve in the image plane. The small brown filled
cross is the predicted position of the source quasar, and the small
orange triangle is the predicted position of the center of the lens
mass.  The differences between the observed and modelled image positions
are much smaller than the observational uncertainties.  The flux ratios
predicted from the model, B/A, C/A and D/A are 0.78, 0.43 and 0.22,
respectively. The total magnification of the quasar images which is
predicted by the model is 56.48. The predicted flux ratios are close to
the observational results; B/A$=0.69{\pm}0.04$, C/A$=0.46{\pm}0.02$, and
D/A$=0.25{\pm}0.01$ (measured from the $i$ band image).  Microlensing by
substructures and/or reddening by the \ion{Mg}{2} absorption line
systems that are seen in each spectrum might be the cause of the
differences between the predicted flux ratios and the observations.
}
\end{document}